\documentclass[prb,amsmath,twocolumn,superscriptaddress,showpacs]{revtex4}
\usepackage{bm}
\usepackage{amssymb}
\usepackage{graphicx}

\def\b0{{\bf{0}}}
\begin{document}
\title{Quantum coherence and entanglement induced by the continuum between distant localized states}
\author{Jing Ping}
\affiliation{State Key Laboratory for Superlattices and Microstructures, Institute of Semiconductors, Chinese Academy of Sciences, P.O. Box 912, Beijing 100083, China}
\author{Xin-Qi Li}
\affiliation{Department of Physics, Beijing Normal University, Beijing 100875, China}
\affiliation{State Key Laboratory for Superlattices and Microstructures, Institute of Semiconductors,
Chinese Academy of Sciences, P.O. Box 912, Beijing 100083, China}
\author{Shmuel Gurvitz}
\email{shmuel.gurvitz@weizmann.ac.il}
\affiliation{Department of Particle Physics and Astrophysics, Weizmann Institute of Science, Rehovot 76100, Israel}

\date{\today}


\begin{abstract}
It is demonstrated that two distant quantum wells separated by a reservoir with a continuous spectrum can possess bound eigenstates embedded in the continuum. These represent a linear superposition of quantum states localized in the wells. We show that such a state can be isolated in the course of free evolution from any initial state by a null-result measurement in the reservoir. The latter might not be necessary in the many-body case. The resulting superposition is regulated by ratio of couplings between the wells and the reservoir. In particular, one can lock the system in one of the wells by enhancing this ratio. By tuning parameters of the quantum wells, many-body entangled states in distant wells can be produced through interactions and statistics.
\end{abstract}

\maketitle

\section{Introduction}
Quantum coherence and entanglement are basic properties of quantum systems that play a crucial role in many phenomena, in particular in those related to quantum information and quantum computing \cite{nielsen}. These properties, however, are usually destroyed by the environment \cite{breuer}. This can happen because of coupling of the quantum systems to infinite reservoirs that generates random fluctuations of the system parameters or direct leakage (decay) to the reservoirs.

Recent works indicate, however, that under certain conditions the environment may endorse the creation of entanglement between two physical systems, like two spins or two oscillators \cite{knight,ent1,wolf}. In this paper we pronounce a positive role of the environment in quantum coherence and entanglement even more. We demonstrate that it can drive the system to a stable superposition of spatially separated quantum states. Moreover such a superposition can be controlled by varying the system's parameters.

\begin{figure}[h]
\includegraphics[width=7cm]{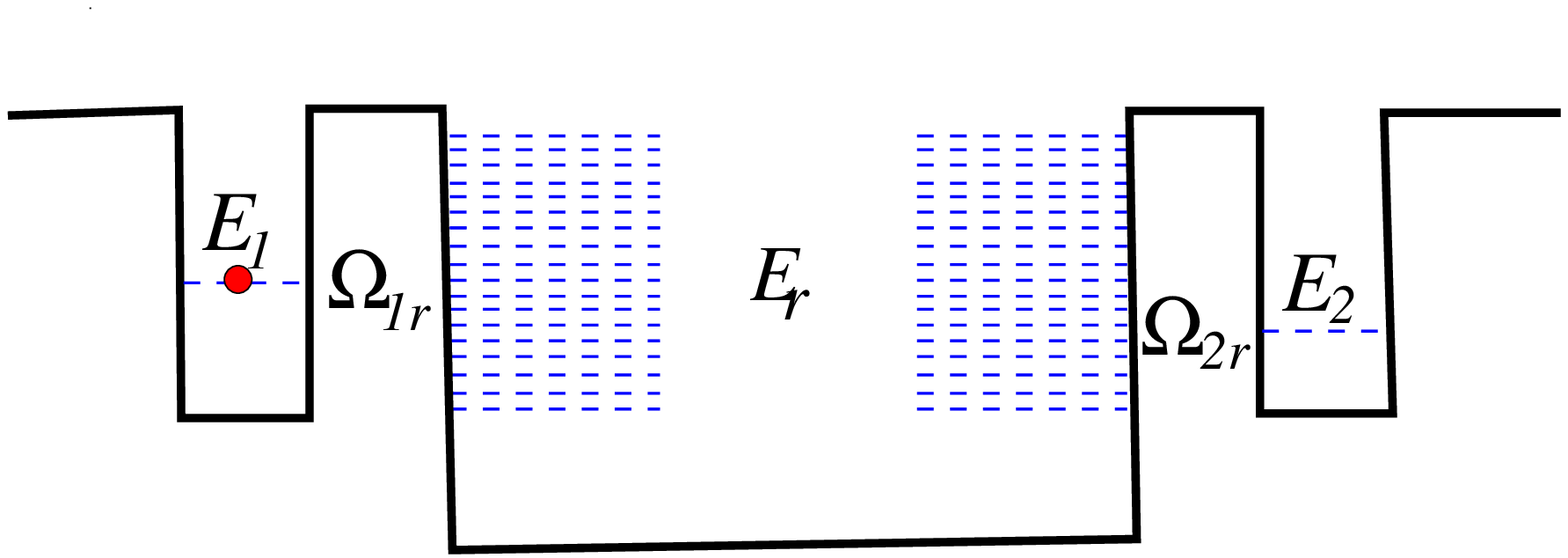}
\caption{(color online) A particle in two quantum wells separated by a reservoir. $E_{1,2}$ are the energy levels of the wells and $\Omega_{1r}$,  ($\Omega_{2r})$ denote the couplings between the level $E_{1(2)}$ and the level $E_r$ in the reservoir.}
\label{fig1}
\end{figure}

Consider for instance a particle in a quantum double-well. It is well-known that if the particle is prepared in one of the wells it displays Rabi oscillations between the wells. The question is what will happen if the wells are coupled not directly but through a reservoir possessing a continuous spectrum, Fig.~\ref{fig1}. It is quite natural to expect that instead of Rabi oscillations one finds no more than the exponential decay of a particle to the reservoir. For this reason the dynamics of a quantum system inside {\em two} wells, separated by a reservoir, has not attracted much attention. \footnote{It was observed in Ref.~\onlinecite{g1}, without any further analysis, that  there is a finite probability of finding a particle inside the dots at $t\to\infty$ in the case of identical wells. }

In this paper we demonstrate that, contrary to expectation, the system of quantum wells separated by, and coupled to, a continuum can reveal very peculiar properties. For instance, one finds localized eigenstates embedded in the continuum. Moreover, in the case of aligned levels, $E_1=E_2$ (Fig.~\ref{fig1}), the system evolves inexorably to such a state. This means that at large times there is a finite probability of finding the system in a localized state inside the wells, even though the wells are well separated. This state is a linear superposition of localized states in different wells. If $E_1\not =E_2$, but their difference is smaller than the level width, the system decays to the continuum in the end. During its evolution, however, the system spends time in the superposition state inside the wells and the lifetime of this state can be very long.

The same phenomenon takes place in the many-body case, displaying additional features. These are the result of particle statistics and of particle interactions. At large times one can find the many-particle system in linear superpositions of states inside the distant wells, which are entangled. The most important properties of such systems is that the superposition and entanglement can be controlled by changing the wells' parameters, such as the energy levels and their coupling with the reservoir. This can make these systems very useful for development of new quantum devices and their applications for quantum information and quantum computing.

The plan of this paper is as follows: In Sec.~\ref{Sec2} we investigate one-particle dynamics in two wells separated by a reservoir. We demonstrate the appearance of stable states localized inside the wells. In Sec.~\ref{Sec3} we introduce the optimal basis for the two-well states by a unitary transformation (resulting in ``dark'' states) whereby the problem is greatly simplified. In this basis the mechanism for appearance of isolated, localized states in the continuum can be easily understood. Also the extension to the many-body case is immediate, including in particular the effects of the Pauli principle and of particle interaction. The many body case is discussed separately for fermions in Sec.~\ref{Sec4} and for bosons in Sec.~\ref{Sec5}. The last section is a summary and discussion.

\section{Particle in quantum wells attached to reservoir \label{Sec2}}

Consider a single particle in two quantum wells separated by a reservoir, Fig.~\ref{fig1}. Such a system can be realized, for instance, by an electron (fermion) in two quantum dots or by a photon (boson) in two separated quantum cavities. We consider a spinless electron. We assume that each of the dots contains only one level ($E_1^{}$ and $E_2^{}$), whereas the reservoir states, $E_r^{}$, are very dense. The system can be described by the following tunneling Hamiltonian,
\begin{align}
H=&E_1^{}a_1^\dagger a_1^{}+E_2^{}a_2^\dagger
a_2^{}+\sum_rE_r^{}a_r^\dagger
a_r^{}\nonumber\\&+\sum_r(\Omega_{1r}^{}a_r^\dagger
a_1^{}+\Omega_{2r}^{}a_r^\dagger a_2^{}+H.c.)\, . \label{a1}
\end{align}
Here $a_{1,2}^{}$ and $a_r^{}$ are annihilation operators of an
electron in the quantum dots or in the reservoir. In the absence of a magnetic field, the couplings $\Omega_{1r}^{}$ and $\Omega_{2r}^{}$ are real, but they can be of opposite sign, depending on the relative parity of the states $E_{1,2}$ in the quantum dots.

The wave function of an electron in this system can be written in the  most general way as \begin{align} |\Psi (t)\rangle
=\big[b_1^{}(t)a_1^\dagger +b_2^{}(t)a_2^\dagger +\sum_r
b_r^{}(t)a_r^\dagger\big]|0\rangle\, , \label{a2}\end{align} where
$b_{1,2}(t)$ and $b_r(t)$ are the probability amplitudes of finding
the electron in the dots or in the reservoir, respectively. These amplitudes are obtained from the Schr\"odinger equation $i\partial_t|\Psi (t)\rangle=H|\Psi (t)\rangle$. For its solution it is useful to apply the Laplace transform, $|\Psi (t)\rangle\to|\tilde\Psi (E)\rangle =\int_0^\infty |\Psi (t)\rangle\exp (iEt)dE$. Then the time dependent Schr\"odinger equation becomes the algebraic equation
\begin{align}
(E-H)|\tilde\Psi (E)\rangle=i|\Psi(0)\rangle\label{a3}
\end{align}
It can be written explicitly as
\begin{subequations}
\label{a4}
\begin{align}
&(E-E_1)\tilde b_1(E)-\sum_r\Omega_{1r} \tilde b_r(E)=ib_1(0)\label{a4a}\\
&(E-E_2)\tilde b_2(E)-\sum_r\Omega_{2r}\tilde b_r(E)=ib_2(0)\label{a4b}\\
&(E-E_r)\tilde b_r(E)-\Omega_{1r} \tilde b_1(E)-\Omega_{2r} \tilde
b_2(E)=ib_r(0)\label{a4c}
\end{align}
\end{subequations}
The r.h.s. of Eqs.~(\ref{a4}) corresponds to the initial conditions.

Let us assume that the electron is initially localized inside the dots,
\begin{align}
|\Psi (0)\rangle =C_1\, a_1^\dagger |0\rangle +C_2\, a_2^\dagger
|0\rangle\, , \label{a5} \end{align}
where $|C_1|^2+|C_2|^2=1$. Then solving Eq.~(\ref{a4c}) for $\tilde b_r(E)$ and substituting the result into Eqs.~(\ref{a4a}), (\ref{a4b}) we
obtain
\begin{subequations}
\label{a7}
\begin{align}
&\left(E-E_1-\sum_r{\Omega_{1r}^2\over E-E_r} \right)\tilde
b_1
-\sum_r{\Omega_{1r}\Omega_{2r}\over E-E_r}\tilde b_2=iC_1\label{a7a}\\
&\left(E-E_2-\sum_r{\Omega_{2r}^2\over E-E_r}
\right)\tilde b_2 -\sum_r{\Omega_{1r}\Omega_{2r}\over E-E_r}\tilde
b_1=iC_2\label{a7b}
\end{align}
\end{subequations}
Since the levels in the reservoir are very dense, we can replace
$\sum_r\to \int\rho (E_r)dE_r$, where $\rho(E_r)$ is the density of
states, and $\Omega_r\to\Omega (E_r)$. Therefore we can evaluate the
sums in Eqs.~(\ref{a7}) as
\begin{align}
\sum_r{\Omega_{jr}\Omega_{j'r}\over E-E_r}\to\int{\Omega_j
(E_r)\Omega_{j'}(E_r)\over E-E_r}\rho (E_r)dE_r={\cal F}_{jj'}(E)\, ,
\end{align}
where $j,j'=1,2$. As a result, Eqs.~(\ref{a7}) can be rewritten as
\begin{subequations}
\label{amp}
\begin{align}
&\left[E-E_1-{\cal F}_{11}(E)\right]\tilde b_1(E)
-{\cal F}_{12}(E)\tilde b_2(E)=iC_1\label{ampa}\\
&\left[E-E_2-{\cal F}_{22}(E)\right]\tilde b_2(E) -{\cal
F}_{12}(E)\tilde b_1(E)=iC_2\label{ampb}
\end{align}
\end{subequations}
Solving these equations we find the amplitudes $\tilde
b_{1,2}(E)$. The corresponding time-dependent amplitudes
$b_{1,2}(t)$ are obtained from the inverse Laplace transform,
\begin{align}
b(t)={1\over 2\pi}\int_{-\infty}^\infty\tilde b(E) e^{-iEt}dE
\label{invlap}
\end{align}
This is an exact solution of the problem.

Let us assume that the density of states and the tunneling couplings are weakly dependent on energy, so that they are taken as constants, $\Omega_j(E_r)\to\Omega_j$ and $\rho (E_r)\to\rho$. Then for the large  cut-off, $\Lambda\to\infty$, we obtain
\begin{align}
{\cal F}_{jj'}(E)=\int_{-\Lambda}^{\Lambda}
{\Omega_j\Omega_{j'}\over
E-E_r}\rho_RdE_r=-i\eta_{jj'}^{}
{\sqrt{\Gamma_j\Gamma_{j'}}\over2}\, ,
\label{f1}
\end{align}
where $\Gamma_j=2\pi\Omega_j^2\rho$ is the width of the level $E_j$ and $\eta_{jj'}^{} =(\Omega_j\Omega_{j'})/|\Omega_j\Omega_{j'}|=\pm 1$ is the relative parity (number of nodes) of the dot states \cite{lap}.

Now we apply the inverse Laplace transform (\ref{invlap}) to Eqs.~(\ref{amp}). Since the coefficients ${\cal F}_{jj'}$, Eq.~(\ref{f1}), are independent of $E$, Eqs.~(\ref{amp}) are transformed to the following linear equations for the amplitudes $b_{1,2}(t)$,
\begin{subequations}
\label{a10}
\begin{align}
&i\dot b_1(t)=\left(E_1-i{\Gamma_1\over2}\right)b_1(t)-
i\eta_{12}^{}{\sqrt{\Gamma_1\Gamma_2}\over2}b_2(t)\label{a10a}\\
&i\dot b_2(t)=\left(E_2-i{\Gamma_2\over2}\right)b_2(t)-
i\eta_{12}^{}{\sqrt{\Gamma_1\Gamma_2}\over2}b_1(t)\label{a10b}
\end{align}
\end{subequations}
Note that $\int_{-\infty}^\infty C_{1,2}\exp (-iEt)dE=2\pi \delta (t)=0$, since $t>0$ in the inverse Laplace transform.

Using these amplitudes, we can evaluate the (reduced) density-matrix, $\sigma_{jj'}(t)$,
defined as
\begin{align}
&\sigma_{11}(t)=|b_1(t)|^2,~~\sigma_{22}(t)=|b_2(t)|^2,
~~\sigma_{12}(t)=b_1(t)b_2^*(t)\nonumber\\&
\sigma_{00}(t)=\sum_r|b_r(t)|^2=1-\sigma_{11}(t)-\sigma_{22}(t)\, ,
\label{dens}
\end{align}
where $\sigma_{00}(t)$ is the probability of finding the electron in the reservoir.

Multiplying Eqs.~(\ref{a10}) by $b_{1,2}^*(t)$ and subtracting its complex
conjugate, one can easily transform these equations to the
following master equations for $\sigma_{jj'}(t)$,
\begin{subequations} \label{a11}
\begin{align}
&\dot\sigma_{11}^{}(t)=-\Gamma_1\sigma_{11}^{}(t)
-\eta_{12}^{}{\sqrt{\Gamma_1\Gamma_2}\over2}[\sigma_{12}^{}(t)+\sigma_{21}^{}(t)]
\label{a11a}\\
&\dot\sigma_{22}^{}(t)=-\Gamma_2\sigma_{22}^{}(t)
-\eta_{12}^{}{\sqrt{\Gamma_1\Gamma_2}\over2}[\sigma_{12}^{}(t)+\sigma_{21}^{}(t)]
\label{a11b}\\
&\dot\sigma_{12}^{}(t)=i(E_2^{}-E_1^{})\sigma_{12}^{}(t)
-\eta_{12}^{}{\sqrt{\Gamma_1\Gamma_2}\over2}[\sigma_{11}^{}(t)+\sigma_{22}^{}(t)]
\nonumber\\
&~~~~~~~~~~~~~~~~~~~~~~~~~~~~~~~~~~~~~~~~~
-{\Gamma_1+\Gamma_2\over2}\sigma_{12}^{}(t)\label{a11c}
\end{align}
\end{subequations}
Similar equations have been considered in the literature for describing electron transport through two dots in sequence, separated by a reservoir \cite{g1}, or through parallel dots \cite{dong,feng}.

Consider now the symmetric case, $\Gamma_1=\Gamma_2=\Gamma$ and $\eta_{12}^{}=1$. Then solving Eqs.~(\ref{a11}) with the initial conditions $\sigma_{11}(0)=1$ and $\sigma_{22}(0)=\sigma_{12}(0)=0$ we find that
\begin{subequations}
\label{a12}
\begin{align}
\sigma_{11}(t)&={\Gamma^2\cosh^2(\omega^{}t/2)-\varepsilon^2\over
\omega^2}e^{-\Gamma t}\, , \label{a12a}\\
\sigma_{22}(t)&={\Gamma^2\sinh^2(\omega^{}t/2)\over
\omega^2}e^{-\Gamma t}\, , \label{a12b}\\
\sigma_{12}(t)&=-\frac{i\varepsilon [1-\cosh
   (\omega t )]+\omega \sinh (\omega t )}{2
   \omega^2}\Gamma e^{-\Gamma t}\, ,\label{a12c}
\end{align}
\end{subequations}
where $\varepsilon =E_1-E_2$ and $\omega
=\sqrt{\Gamma^2-\varepsilon^2}$.

It follows from Eqs.~(\ref{a12}) that, as expected,
the probability of finding the electron in the dots vanishes in the asymptotic limit, $\sigma_{jj'}(t)\to 0$ as $e^{-(\Gamma-\sqrt{\Gamma^2-\varepsilon^2})\, t}$ for any $\varepsilon\not =0$. If $\varepsilon =0$, however, then $\sigma_{11}(t)\to 1/4$, $\sigma_{22}(t)\to 1/4$ and $\sigma_{12}(t)\to -1/4$ for $t\to\infty$. Therefore one finds  the electron localized in a linear superposition inside the dots with  probability $1/2$, despite the coupling of this state to the continuous spectrum of the reservoir.

The non-analyticity of the density matrix as a function of $\varepsilon$ appears only in the asymptotic limit. At any finite $t$ the density matrix is an analytic function of $\varepsilon$. Indeed, the probability of finding the electron inside the dots at finite $t$ is not zero, although it is negligibly small for large $t$. However, the dwell-time of the electron inside the dots ($\tau$) grows as $\varepsilon$ decreases. By expanding the slowest exponent in Eqs.~(\ref{a12}) in powers $\varepsilon /\Gamma$ one easily finds that  $\tau\simeq 2\Gamma/\varepsilon^2$. Thus the electron occupies the dots for a long time when $\varepsilon\lesssim \Gamma$. This dwell-time diverges when $\varepsilon\to 0$. Such an unexpected localization effect in a continuous spectrum is a manifestation of quantum interference. As we shall see, this can be understood in a much more transparent way if we use a different basis for the electron states in the two dots.

\section{Optimal basis \label{Sec3}}

Consider a unitary transformation from the dot states to a different basis,  $a_{1,2}^\dagger |0\rangle\to c_{1,2}^\dagger |0\rangle$, where
\begin{subequations}
\label{un}
\begin{align}
a_1^{}&=\cos\alpha\, c_1^{}+\sin\alpha\, c_2^{} \label{una}\\
a_2^{}&=-\sin\alpha\, c_1^{}+\cos\alpha\, c_2^{} \label{unb}
\end{align}
\end{subequations}
In this new basis the Hamiltonian (\ref{a1}) reads
\begin{align}
&H=E'_1c_1^\dagger c_1^{}+E'_2c_2^\dagger
c_2^{}+{1\over2}\varepsilon\sin 2\alpha\,
(c_1^{\dagger}c_2^{}+c_2^\dagger c_1^{})
\nonumber\\&+\sum_rE_ra_r^\dagger a_r^{}+\sum_r(g_1^{}a_r^\dagger
c_1^{}+g_2^{}a_r^\dagger c_2^{}+H.c.)\, , \label{ham}
\end{align}
where
$E'_{1,2}=E_{1,2}^{}\cos^2\alpha+E_{2,1}^{}\sin^2\alpha$
and
\begin{subequations}
\label{ncoup}
\begin{align}
g_1^{}&=\Omega_1\cos\alpha-\Omega_2\sin\alpha\label{ncoup1}\\
g_2^{}&=\Omega_1\sin\alpha+\Omega_2\cos\alpha\label{ncoup2}
\end{align}
\end{subequations}

\begin{figure}[h]
\includegraphics[width=7cm]{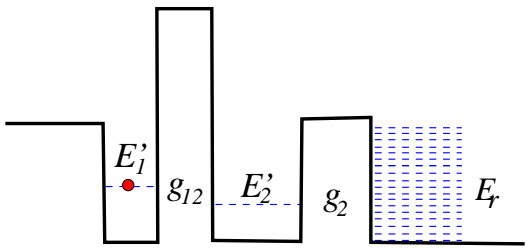}
\caption{(color online) Double-dot system corresponding to the Hamiltonian (\ref{ham}), representing the separated dots in the new basis, $c_{1,2}^\dagger |0\rangle$.} \label{fig3}
\end{figure}

Let us choose the new basis such that one of the states $c_{1,2}^\dagger |0\rangle$ is decoupled from the reservoir \cite{feng} (``dark'' state). This can be realized for $\tan \alpha =\Omega_1/\Omega_2$, resulting in $g_1=0$. Then one finds from Eq.~(\ref{ham}) that the original separated-dot system, Fig.~\ref{fig1}, is mapped to a double-dot system coupled to one reservoir, Fig.~\ref{fig3}, where the inter-dot coupling between the states $c_{1,2}^\dagger |0\rangle$ is given by
\begin{align}
g_{12}^{}=\varepsilon{\Omega_1\Omega_2\over \Omega_1^2+\Omega_2^2}\, .
\label{intcoup}
\end{align}
Thus in the case of aligned levels, $\varepsilon =0$, the inter-dot coupling vanishes, $g_{12}^{}=0$, leading to total decoupling of the state $c_1^\dagger |0\rangle$ from the reservoir, similar to a dark state in quantum optics and atomic physics (see for instance Ref.~\onlinecite{plenio}). Therefore this state, which is a linear superposition in the original basis,
\begin{align}
c_1^\dagger |0\rangle =\big(\cos\bar\alpha\, a_1^\dagger -\sin\bar\alpha\, a_2^\dagger\big )|0\rangle \, ,\label{asstate}
\end{align}
where $\cos\bar\alpha =\Omega_2^{}/\sqrt{\Omega_1^2+\Omega_2^2}$, survives in the asymptotic limit $t\to\infty$ with probability
$P_0=|\langle 0|c_1^{}\Psi (0)\rangle|^2$.

On the other hand, the state $c_2^\dagger |0\rangle$ decays to the reservoir with a rate
\begin{align}
\Gamma'_2=2\pi g_2^2\rho=\Gamma_1+\Gamma_2\, .
\label{rate2}
\end{align}
The probability of finding the electron in the reservoir at $t\to\infty$ is therefore $P_1=|\langle 0|c_2^{}\Psi (0)\rangle |^2 =1-P_0$. For the initial condition corresponding to the electron in the linear superposition of the two-dot states, Eq.~(\ref{a5}), we obtain \footnote{We do not consider a case where the electron is initially in a superposition between the reservoir and one of the wells, since preparation of such a state would be extremely complicated.}
\begin{align}
P_{0}={\big|C_1^{}\eta_{12}^{}\sqrt{y}-C_2^{}\big|^2\over 1+y},~~P_{1}={\big|C_1^{}+C_2^{}\eta_{12}^{}\sqrt{y}\big|^2\over 1+y}\, ,
\label{prob}
\end{align}
where $y=\Gamma_2/\Gamma_1$.

It follows from Eq.~(\ref{prob}) that if the electron is initially in the superposition (\ref{asstate}), i.e. $C_1=\eta_{12}^{}\sqrt{y}/\sqrt{1+y}$ and $C_2=-1/\sqrt{1+y}$,
the probability of finding it in the reservoir at $t\to\infty$ is $P_1=0$. Thus the state $c_1^\dagger |0\rangle$ offers a striking example of a bound state embedded in the continuum \cite{emb1}, which has not been discussed in the literature to our knowledge.

In fact, in order to prepare such a bound state embedded in the continuum we do not need any special initial conditions. One can start with any initial condition, for instance, placing an electron in one of the dots in Fig.~\ref{fig1}. Then by selecting states where {\em no} electrons are found in the reservoir (a ``null measurement'') for $t\gg 1/\Gamma'_2$, we project the system to the state (\ref{asstate}).

Consider now the electron density matrix Eq.~(\ref{dens}) in the asymptotic limit, $\bar\sigma_{jj'}=\sigma_{jj'}(t\to\infty )$. For the initial conditions corresponding to having the left-hand dot in Fig.~\ref{fig1} occupied, one easily obtains from Eqs.~(\ref{asstate}), (\ref{prob})
\begin{align}
&\bar\sigma_{11}^{}={y^2\over (1+y)^2},~~~~
\bar\sigma_{22}^{}={y\over (1+y)^2},\nonumber\\
&\bar\sigma_{12}^{}=-{y^{3/2}\over (1+y)^2},~~~~\bar\sigma_{00}^{}={1\over 1+y}\, .
\label{prob2}
\end{align}
It follows from these expressions that one can influence the occupation of the dots in the asymptotic limit by varying the barriers'  penetrability. For instance, the occupation of the left dot, $\bar\sigma_{11}^{}$,  increases with $y$, so that $\bar\sigma_{11}^{}\to 1$ (and respectively $\bar\sigma_{00}^{}\to 0$) when $y\to\infty$. This tells us that, contrary to expectation, localization of the electron increases when the right barrier becomes more transparent. It resembles the increase of dwell time of a two-level system coupled to the continuum when the coupling to the continuum is increased \cite{bar,bar1}. Here, however the phenomenon is more surprising, since the dots are far from each other and the localization time is infinite.

\begin{figure}[h]
\includegraphics[width=7cm]{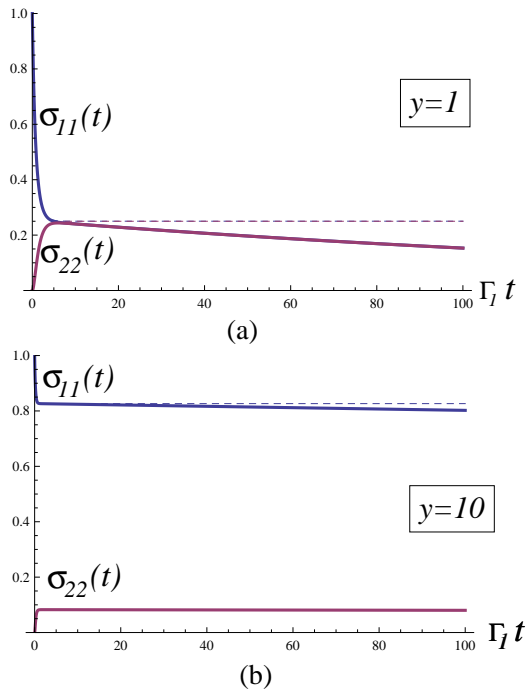}
\caption{(color online) Occupation probabilities of the first and the second dots of Fig.~\ref{fig1}, $\sigma_{11}(t)$ and $\sigma_{22}(t)$, as a function of time for (a) $y=1$ (symmetric dots) and (b) $y=10$ (asymmetric dots). Dashed lines correspond to $\varepsilon =0$ and solid lines to $\varepsilon /\Gamma_1 =0.1$. }
\label{fig2}
\end{figure}

It follows from Eq.~(\ref{intcoup}) that the state $c_1^\dagger |0\rangle$ is not  isolated if $\varepsilon\not =0$. In this case the electron eventually decays to the reservoir. Nevertheless, its dwell time $\tau$ inside the dots can be very large for small $\varepsilon\not =0$. It corresponds to the decay time for an electron placed in the inner dot of Fig.~\ref{fig3}. This quantity can be easily evaluated by using the results of Refs.~[\onlinecite{gmar,bar}]. One finds for the electron initially localized in the left-hand dot of Fig.~\ref{fig1},
\begin{align}
\tau={\Gamma'_2\over 4 g_{12}^2}={\Gamma_1\over \varepsilon^2}{(1+y)^3\over
4 y}\, .
\label{tau}
\end{align}
Therefore, in the non-symmetric case $y\not =1$, the dwell time increases as $y^2$ for large $y$ or as $1/y$ for small $y$. This implies that the localization of an electron inside the dots is more pronounced for the asymmetric case and therefore it does not require a precise tuning of the levels as in the symmetric case $y=1$. This is demonstrated in Fig.~\ref{fig2}, where the time-dependence of the occupation probabilities given by Eqs.~(\ref{a12}) is shown for several choices of $\varepsilon$ and $y$.

We have shown that the asymptotic state of a quantum particle in separated wells can be analyzed without any detailed calculations such as those in Sec.~\ref{Sec2}. All we need is the unitary transformation (\ref{un}) of the Hamiltonian (\ref{a1}), leading to a dark state---the optimal basis. The possibility of performing such a transformation is not restricted by the assumptions made in Sec.~\ref{Sec2} regarding the reservoir spectrum and the energy dependence of the coupling amplitudes. One needs only that the couplings $\Omega_{1(2)r}^{}$ in the Hamiltonian (\ref{a1}) satisfy the following condition:
\begin{align}
{\Omega_{2r}^{}\over\Omega_{1r}^{}}=\sqrt{y}={\mbox{\rm const}}.
\label{cond}
\end{align}
Then our results remain largely unchanged.

\section{Many-electron case \label{Sec4}}
Now we are going to analyze many-body localized states separated by a continuum. We use the same optimal basis, which greatly simplifies the treatment also in this case and makes it transparent.
\subsection{Two electrons in separated dots}

Consider first two spinless electrons inside the separated dots in Fig.~\ref{fig1}. We neglect the electron--electron interaction between the dots due to their large separation. Also, we do not need to include explicitly the intra-dot electron--electron interaction since two spinless electrons cannot occupy the same dot. Therefore we can describe the entire system by the same non-interacting Hamiltonian (\ref{a1}) that we used before.

The initial wave function of two electrons occupying the two dots can only be $|\Psi (0)\rangle =a_1^\dagger a_2^\dagger |0\rangle$. Let us rewrite it in the rotated basis, Eq.~(\ref{un}). We obtain
\begin{align}
|\Psi (0)\rangle &=(\cos\bar\alpha\, c_1^\dagger+\sin\bar\alpha\, c_2^\dagger )
(-\sin\bar\alpha\, c_1^\dagger+\cos\bar\alpha\, c_2^\dagger )\nonumber\\
&=c_1^\dagger c_2^\dagger |0\rangle\, .
\label{2el}
\end{align}
Thus both wells in Fig.~\ref{fig3} are initially occupied. Consider the case of aligned levels, $\varepsilon =0$. In this case the left well in Fig.~\ref{fig3} is decoupled from the right well. Then the asymptotic state of the entire system can be easily determined: the state $c_2^\dagger |0\rangle$ decays to the continuum, whereas the state $c_1^\dagger |0\rangle$
remains in the left well as $t\to\infty$. Thus the final state of the system corresponds to one electron in the continuum and the second electron inside the dots in the linear superposition (\ref{asstate}). The corresponding density matrix in the asymptotic limit, $\bar\sigma_{jj'}^{(m)}$, where $m=0,1,2$ denotes the number of electrons in the reservoir, is given by
\begin{align}
&\bar\sigma_{11}^{(1)}={y\over 1+y},~~~~~~
\bar\sigma_{22}^{(1)}={1\over 1+y},\nonumber\\
&\bar\sigma_{12}^{(1)}=-{\sqrt{y}\over 1+y},~~~~\bar\sigma_{jj'}^{(0)}=0,
~~~~\bar\sigma_{00}^{(2)}=0\, .
\label{2el1}
\end{align}
Note that, in contrast with Eq.~(\ref{prob2}), the asymptotic probability of finding one electron inside the two dots is one. Thus the bound state in the continuum can be prepared without any selective measurements---one needs only full initial occupation of the  separated dots. Then the Schr\"odinger evolution alone drives the system to the asymptotic state with one of the electrons localized inside the distant quantum dots.

\subsection{Parallel dots separated by the reservoir}

\begin{figure}[h]
\includegraphics[width=7cm]{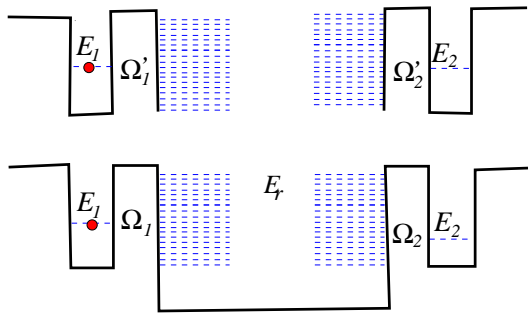}
\caption{(color online) Two electrons in parallel quantum dots separated by a common reservoir.}
\label{fig4}
\end{figure}

In order to realize a controlled entangled state of two interacting electrons, we consider the generic setup shown in Fig.~\ref{fig4}. It represents two pairs of parallel dots separated by a common reservoir, where two spinless electrons occupying the parallel dots can stay in close proximity. For simplicity we assume that that $\Omega'_1/\Omega_1=\Omega'_2/\Omega_2=y'$. The  Hamiltonian describing the entire system can be written as
\begin{align}
H&=E_1^{}(a_1^\dagger a_1^{}+a_1^{\prime\dagger} a_1^{\prime})+E_2^{}(a_2^\dagger
a_2^{}+a_2^{\prime\dagger}a_2^{\prime})
+\sum_rE_ra_r^\dagger
a_r^{}\nonumber\\&+\sum_r[\Omega_1^{}a_r^\dagger
(a_1^{}+y'a_1^\prime)+\Omega_2^{}a_r^\dagger (a_2^{}+y' a_2^\prime)+H.c.]\nonumber\\
&+U(a_1^\dagger a_1^{}a_1^{\prime\dagger} a_1^{\prime}+
a_2^\dagger a_2^{}a_2^{\prime\dagger}a_2^{\prime})
\,. \label{ham1}
\end{align}
Here $a_{1,2}^{\prime}$ are the annihilation operators of
electrons in the upper quantum dots, whereas $a_{1,2}^{}$ refer to the lower dots. The last term describes the Coulomb repulsion between two electrons in parallel dots.

We first apply the unitary transformation (\ref{un}) to the left and the right parallel dots,
\begin{subequations}
\label{unp}
\begin{align}
a_{j}^{}&=\cos\bar\beta\, d_{j}^{}+\sin\bar\beta\, d_{j}^{\prime } \label{unpa}\\
a_{j}^{\prime }&=-\sin\bar\beta\, d_{j}^{}+ \cos\bar\beta\, d_{j}^{\prime } \label{unpb}
\end{align}
\end{subequations}
where $\cos\bar\beta =y'/\sqrt{1+y^{\prime\, 2}}$ and $j=1,2$. In these variables the Hamiltonian (\ref{ham1}) reads
\begin{align}
&H=E_1^{}(n_1^{}+n_1^{\prime})+E_2^{}(n_2^{}
+n_2^{\prime})+U(n_1^{}n_1^\prime +n_2^{}n_2^\prime)
\nonumber\\&+\sum_rE_ra_r^\dagger
a_r^{}+\sum_r\sqrt{1+y^{\prime\, 2}}(\Omega_1^{}a_r^\dagger
d_1^\prime+\Omega_2^{}a_r^\dagger d_2^\prime+H.c.)
\label{ham2}
\end{align}
where $n_{j}^{}=d_{j}^\dagger d_{j}^{}$ and $n_{j}^{\prime}=d_{j}^{\prime\dagger} d_{j}^{\prime}$.
Note that the electron--electron interaction term is invariant under the unitary transformation (\ref{unp}).

\begin{figure}[h]
\includegraphics[width=7cm]{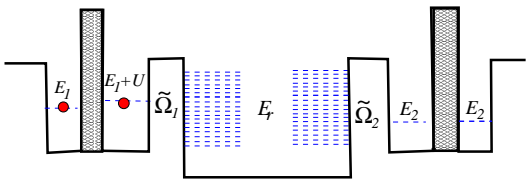}
\caption{(color online) Coupled-dot system represented by the Hamiltonian (\ref{ham2}) in the rotated basis (\ref{unp}). The two outermost dots are decoupled from the inner dots. $\tilde\Omega_{1,2}^{}=\sqrt{1+y^{\prime\, 2}}\Omega_{1,2}^{}$ are the couplings of the inner dots with the reservoir.}
\label{fig5}
\end{figure}

It follows from Eq.~(\ref{ham2}) that the states $d^\dagger_{1,2}|0\rangle$ are decoupled from the reservoir, so the operator $n_j^{}$ commutes with the Hamiltonian. As a result $n_j^{}$ can be replaced by its eigenvalue, $n_j^{}\to \bar n_j^{}=0,1$, corresponding to the initial state of the system. Then the problem becomes equivalent to that of one electron in the separated dots, Fig.~\ref{fig1}, described by the Hamiltonian (\ref{a1}), with $E_{j}\to E_{j}+U \bar n_{j}^{}$ and $a_{j}^{}\to d_{j}^{\prime}$. Thus the system described by the Hamiltonian (\ref{ham1}) is mapped to two coupled-dot systems, separated by the reservoir, where the outer dots are decoupled from the inner dots, as shown in Fig.~\ref{fig5}.

Consider for instance the time-evolution of two electrons occupying the two dots on the left in  Fig.~\ref{fig4} at $t=0$. The initial state wave function is (cf\@. Eq.~(\ref{2el})),
\begin{align}
|\Psi (0)\rangle =a_1^\dagger a_1^{\prime\,\dagger}|0\rangle =d_1^\dagger d_1^{\prime\,\dagger}|0\rangle
\label{wf2}
\end{align}
This corresponds to having the left dots in Fig.~\ref{fig5} occupied.
Then, if $E_2\not =E_1+U$, the electron in the inner dot decays to the continuum. As a result, in the asymptotic state we find one electron in the linear superposition $d_1^\dagger |0\rangle=(1/\sqrt{1+y^{\prime\, 2}})(y'a_1^\dagger-a_1^{\prime\,\dagger})|0\rangle$. If, however, we tune the energy levels such that $E_2 =E_1+U$, then both electrons can be found inside the dots in the superposition $(1/\sqrt{1+y^2})d_1^\dagger(y\, d_1^{\prime\,\dagger}- d_2^{\prime\,\dagger})|0\rangle$ (cf\@. Eq.~(\ref{asstate})), with probability $\Gamma_2/(\Gamma_1+\Gamma_2)$ as in Eq.~(\ref{prob}). This example represents a multi-electron bound state embedded in the continuum in the case with electron--electron interaction. This can be reached be selecting states where no electrons are found in the reservoir, similar to a previous case of one electron in two separated wells.

Again the superposition state of two electrons can be regulated by varying the dots' parameters, $y$ and $y'$. For instance, if $y=y'=1$ this asymptotic state reads
\begin{align}
|\Psi_2^{}(t\to\infty )\rangle ={1\over\sqrt{2}}\big[a_1^{}a_1^{\prime\,\dagger}-{1\over2}
(a_1^{}-a_1^{\prime\,\dagger})(a_2^{}+a_2^{\prime\,\dagger})\big]|0\rangle
\label{wf3}\end{align}
This cannot be written as a product of single particle distribution, and therefore it represents an entangled state. However, in the limit where $(y$ or $y')\ll 1$, or $\gg 1$ this state becomes a product, so that the entanglement disappears.

\subsection{Multi-electron entangled states in parallel dots}

Let us consider four electron in the parallel dots, Fig.~\ref{fig4}.
The initial wave function can be written
\begin{align}
|\Psi (0)\rangle =a_1^\dagger a_1^{\prime\,\dagger}a_2^\dagger a_2^{\prime\,\dagger}|0\rangle =d_1^\dagger d_1^{\prime\,\dagger}d_2^\dagger d_2^{\prime\,\dagger}|0\rangle
\label{wf4}
\end{align}
That means thai in the new basis, Eq.~(\ref{unp}), each of the wells is filled by the electron, Fig.~\ref{fig5}. Then the problem is reduced to that of two electrons in the separated dots, Eqs.~(\ref{2el}), (\ref{2el1}). The final state asymptotic state corresponds to one electron in the reservoir whereas the other three electrons are inside the dots in the state
\begin{align}
|\Psi_3^{}(t\to\infty )\rangle &= d_1^\dagger (\cos\bar\alpha\, d_1^{\prime\, \dagger}-\sin\bar\alpha\,d_2^{\prime\, \dagger})d_2^\dagger |0\rangle\nonumber\\[5pt]
&= {y\over\sqrt{1+y^2}}(y\, a_1^\dagger a_1^{\prime\, \dagger}d_2^\dagger +a_2^\dagger a_2^{\prime\, \dagger}d_1^\dagger ) |0\rangle\, ,
\label{wf41}
\end{align}
where
\begin{align}
d_{1,2}^\dagger ={1\over\sqrt{1+y^{\prime\, 2}}}(y'\, a_{1,2}^\dagger -a_{1,2}^{\prime\,\dagger})
\label{wf42}
\end{align}

Equation (\ref{wf41}) describes 3-body entangled state, which is regulated by two parameters, $y$ and $y'$. Let us take for instance $y=y'=1$. Then
\begin{align}
\Psi_3^{}(t\to\infty )= {1\over 4}\Big[a_1^\dagger a_1^{\prime\, \dagger}(a_2^\dagger -a_2^{\prime\, \dagger})
+a_2^\dagger a_2^{\prime\, \dagger}(a_1^\dagger -a_1^{\prime\, \dagger})\Big ] |0\rangle\, ,
\label{wf43}
\end{align}
This state entangles two electrons occupying one pair of the dots with the third electron in the linear superposition of the second pair of dots, Fig.~\ref{fig4}. Note that the Pauli principle plays an important role in this entanglement state by preventing occupation of the same state by two electrons. (Indeed the particle indistinguishability can create entanglement even in the absence of interaction \cite{yaron}). As in the case of two electron in two separated dots, Sec.~\ref{Sec4}A, this state is reached in the course of Schr\"odinger evolution, without any selective measurements in the final state. The same as in the previous case, the state (\ref{wf41}) becomes a product (no entanglement) when $y$ or $y'\ll 1$, or $\gg 1$.

\section{Many-boson states \label{Sec5}}

Obviously, the above results for one fermion (electron) in separated wells remain valid if a fermion is replaced by a  boson. The difference appears in the case of many-bosons. Consider for instance two bosons, each one occupying different dots in Fig.~\ref{fig1}. The wave function describing the the initial state looks the same as for two electrons, $|\Psi (0)\rangle=a_1^\dagger a_2^\dagger|0\rangle$, where $a^\dagger_{1,2}$ are the boson creation operators. Applying the unitary transformation Eq.~(\ref{un})
we rewrite the initial-state wave function as
\begin{align}
|\Psi (0)\rangle =[\sin\bar\alpha\cos\bar\alpha (c_2^{\dagger 2}-c_1^{\dagger 2})+\cos 2\bar\alpha\, c_1^\dagger c_2^\dagger]|0\rangle\, ,
\label{2bos}
\end{align}
It is different from Eq.~(\ref{2el}) for two electrons, which cannot occupy the same state.

As in the previous case, the state $c_2^\dagger |0\rangle$
decays to continuum, Fig.~\ref{fig3}. Thus the probabilities $P_m$ of finding $m$ bosons in the reservoir at $t\to\infty$ are:
\begin{align}
P_0=P_2={2y\over (1+y)^2},~~~P_1={(1-y)^2\over (1+y)^2}\, .
\label{2bos1}
\end{align}
The bosons inside the wells are in the state $c_1^\dagger |0\rangle$ with one boson emitted, or in the state ${1\over\sqrt{2!}}c_1^{\dagger\, 2}|0\rangle$ if no bosons are emitted.
The latter can be rewritten in the initial basis $a_{1,2}^\dagger |0\rangle$, Fig.~\ref{fig1} as
\begin{align}
{1\over 1+y}\big(y\, |2,0\rangle -\sqrt{2\, y}|1,1\rangle +|0,2\rangle \big)\, ,
\label{bwf1}
\end{align}
where $|2,0\rangle$ and  $|0,2\rangle$ denote the states ${1\over \sqrt{2}}a_{1(2)}^{\dagger\, 2}|0\rangle$ corresponding to two bosons in the left (right) well, and $|1,1\rangle$ denotes the state $a_1^\dagger a_2^\dagger |0\rangle$ corresponding to one boson in each well.  The probabilities for these states are respectively, $y^2/(1+y)^2$, $1/(1+y)^2$ and $2y/(1+y)^2$.

The distributions $P_m$, given by Eq.~(\ref{2bos1}) are drastically different from the case of two fermions occupying the dots. In particular, for identical dots, $y=1$, one finds that $P_1=0$. Therefore in this case one boson cannot be emitted to continuum. However, for asymmetric dots in the limit of $y\to 0$ or $y\to\infty$ the asymptotic distributions are the same as in the case of two fermions: one boson is emitted to the reservoir and the another one is localized in one of the dots.

In general case with the initial state corresponding to $N_1$ occupying the left well and $N_2$ bosons occupying the right well in Fig.~\ref{fig1}, the wave function is
\begin{align}
|\Psi(0)\rangle ={1\over\sqrt{N_1!N_2!}}\, a_1^{\dagger\, N_1} a_2^{\dagger\, N_2}|0\rangle\equiv |N_1,N_2\rangle \, .
\label{nboswf}
\end{align}
Using the unitary transformation Eq.~(\ref{un}) with $\cos\alpha =y/\sqrt{1+y^2}$ we obtain
\begin{align}
|\Psi(0)\rangle ={1\over \sqrt{N_1!N_2!}}{(\sqrt{y}\,c_1^\dagger+c_2^\dagger )^{N_1}\over (1+y)^{N_1/2}}{(-c_1^\dagger+\sqrt{y}c_2^\dagger )^{N_2}\over (1+y)^{N_2/2}}|0\rangle
\label{nbos}
\end{align}

Consider first the case where all electrons are initially in the left well, $N_1=N$ and $N_2=0$. Then Eq.~(\ref{nbos}) can be rewritten as
\begin{align}
|\Psi(0)\rangle =\sum_{m=0}^N{y^{(N-m)/2}\over (1+y)^{N/2}}{\sqrt{N!}\over (N-m)!m!}
c_1^{\dagger\, (N-m)}c_2^{\dagger\, m}|0\rangle
\label{nbos1}
\end{align}
Each term of this sum represents a product of the state ${1\over\sqrt{(N-m)!}}c_2^{\dagger\, (N-m)} |0\rangle$ for $N-m$ bosons in the left (inner) well of Fig.~\ref{fig3} and the state
${1\over\sqrt{m!}}c_2^{\dagger\, m} |0\rangle$ for $m$ bosons in the right (outer) well. The latter decays to the reservoir, Fig.~\ref{fig3}, and the former remains inside the dots. Therefore the probability of finding $m$ bosons in the reservoir is
\begin{align}
P_m={y^{N-m}\over (1+y)^{N}}{N!\over (N-m)!m!}\simeq {2^{N+{1\over2}}y^{N-m}\over(1+y)^N\sqrt{\pi N}}e^{-{(N-2m)^2\over 2N}}
\label{nbos2}
\end{align}
It represents a binomial distribution $P_p(m|N)$ where $p=1/(1+y)$. In the strongly asymmetric case, $y\to\infty$, the probability of finding bosons in the reservoir vanishes,
whereas $N$ bosons remain localized in their initial state.

If the bosons are equally distributed between two wells, $N_1=N_2=N$ and the wells are identical, $y=1$, we obtain from Eq.~(\ref{nbos})
\begin{align}
|\Psi(0)\rangle =&{1\over 2^N\,N!}(c_2^{\dagger\,2}-c_1^{\dagger\,2})^N|0\rangle
\nonumber\\[5pt]
&=\sum_{m=1}^N{(-1)^{N-m}
\over 2^N(N-m)!m!}c_1^{\dagger\,2(N-m)}c_2^{\dagger\,2m}|0\rangle
\label{nbos3}
\end{align}
It follows from this equation that only even number of bosons ($2m$) can be found in the reservoir at $t\to\infty$ with probability
\begin{align}
P_{2m}={[2(N-m)]!(2m)!\over 2^{2N}[(N-m)!m!]^2}\simeq {1\over\pi\sqrt{(N-m)m}}\, ,
\label{nbos4}
\end{align}
where we have used the Sterling formula $K!\simeq\sqrt{2\pi K}K^K\exp (-K)$ to evaluate the factorials. It yields flat dependence on $m$ in comparison with the previous case, Eq.~(\ref{nbos2}), when all boson are initially inside one of the dots.

With respect to the remaining $\tilde N=2(N-m)$ bosons inside the wells, these are in the state
\begin{align}{1\over\sqrt{\tilde N!}}c_1^{\dagger\,\tilde N}|0\rangle &={1\over\sqrt{2^{\tilde N}\tilde N!}}(a_1^\dagger -a_2^\dagger)^{\tilde N}|0\rangle\\
&=\sum_{k}{\sqrt{\tilde N!}\over \sqrt{2^{\tilde N}(\tilde N-k)!k!}}|\tilde N-k,k\rangle\, .
\label{nbos5}
\end{align}
Therefore, the probability of finding $k$ bosons in the left well
is given by the binomial distribution $P_{1/2}(k|\tilde N)$ (Eq.~(\ref{nbos2})).

\section{Summary}

In this paper we study behavior of particles (like photons or electrons) in distant quantum wells separated by a common reservoir. Although an entire system possesses a continuum spectrum, we found there localized (bound) states embedded in the continuum. These states represent linear superposition of localized states inside distant wells and manifest large-scale quantum interference phenomenon. They also appear in the multi-particle case in the presence of interaction.

It is quite  remarkable that the bound states embedded in the continuum can be reached asymptotically through the unitary evolution from any initial state. Finally, these state can be selected out by a single null-result measurement in the reservoir, or even without such a selective measurement in many-particle case. One requires only precise alignment of the the energy-levels in different wells, amended by the interaction energy between the particles. If the energy-level alignment is not precise, but the error is smaller than the level width, the system still reaches the superposition state. However, it remains in this state only for a finite time, decaying eventually to the reservoir. Nevertheless, the corresponding dwell-time can be any long, providing that the ratio between the level misalignment and the level-width is arbitrary small.

We found that the superposition is controlled by a relative coupling between different wells and the reservoir. In particular, one can lock a particle in one of the wells by increasing the coupling of the second well with the reservoir. This phenomenon
resembles the Quantum Zeno effect, although no continuous observation is involved. We also found that such an induced localization by a distant  empty well is less dependent on the level alignment. Therefore it is even more accessible to experimental realization than the superposition state in distant wells for a symmetric case.

All our findings are based on the exact treatment of the problem without any weak coupling or Markovian approximations. The central point in our treatment is transition to a new basis containing a ``dark'' state. Then the entire problem becomes very transparent, so that the main features of the superposition state can be obtained straightforwardly without any detailed calculations. In fact, the very existence of the bound state imbedded in the continuum is relied upon the availability of such a basis. This can be establish from general symmetry properties of the Hamiltonian. These are not very restrictive, so that the stable superposition of distant states embedded in the continuum can be found in many different quantum systems.

The optimal basis allows us to treat the many-body case of interacting particles without any essential complications. Then the localized superposition in distant well may include many particle state. These are usually in an entangled state that can be controlled by varying the wells parameters, as well.

We expect that a peculiar controlled superposition of distant localized states can be realized in various quantum systems, like electrons in separated dots, cold atoms in distant traps and photons is quantum cavities. We can also anticipate a possible applicability of such systems to quantum computation as a source of many-particle  distant entangled state.

\begin{acknowledgments} One of us (S.G.) acknowledges Department of Physics, Beijing Normal University, and State Key Laboratory for Superlattices and Microstructures, Institute of Semiconductors, Chinese Academy of Sciences for supporting his visit by the NNSF of China under grants No. 101202101 \&
 10874176, where a part of this work was done. He also thanks M. Heiblum, Y. Silberberg, N. Davidson and B. Svetitsky for useful discussions and important suggestions to this paper.
\end{acknowledgments}

\end{document}